# Marangoni-Induced Reversal of Meniscus-Climbing Microdroplets

Jianxing Sun[1], Patricia B. Weisensee[1,2,*]

[1]*Department of Mechanical Engineering & Materials Science, Washington University in St. Louis, St. Louis, Missouri 63130, USA*

[2]*Institute of Materials Science and Engineering, Washington University in St. Louis, St. Louis, Missouri 63130, USA*

*\* Corresponding author: p.weisensee@wustl.edu*

**Abstract**  Small water droplets or particles located at an oil meniscus typically climb the meniscus due to unbalanced capillary forces. Here, we introduce a size-dependent reversal of this meniscus-climbing behavior, where upon cooling of the underlying substrate, droplets of different sizes concurrently ascend and descend the meniscus. We show that microscopic Marangoni convection cells within the oil meniscus are responsible for this phenomenon. While dynamics of relatively larger water microdroplets are still dominated by unbalanced capillary forces and hence ascend the meniscus, smaller droplets are carried by the surface flow and consequently descend the meniscus. We further demonstrate that the magnitude and direction of the convection cells depend on the meniscus geometry and the substrate temperature and introduce a modified Marangoni number that well predicts their strength. Our findings provide a new approach to manipulating droplets on a liquid meniscus that could have applications in material self-assembly, biological sensitive sensing and testing, or phase change heat transfer.

**Keywords**: Meniscus-Climbing, Droplet Manipulation, Marangoni Convection, Liquid-Infused Surface, Lubricant-Infused Surface (LIS), Capillary Interactions

# 1. Introduction

Liquid menisci naturally develop around wetting objects that protrude liquid surfaces due to capillary forces[1,2]. Insects, such as mesovelia and beetle larva, can locally deform water menisci by modulating their body postures and thus generate a capillary thrust to propel themselves up and down these menisci[3,4]. Meniscus-mediated behaviors of droplets or particles can also easily be observed in our daily life, such as the cheerios effect, where cereals tend to aggregate and form rafts[5], bubble clusters near the walls of soda-filled glasses, and oil drops floating on a bowl of water that migrate away from the solid wall[6]. Meniscus-climbing mechanisms have also been discovered and used for micro/nano-particle self-assembly[7] and enhanced water harvesting[8–11]. Manipulating small objects at liquid interfaces is furthermore becoming increasingly important in areas of microfluidics[12–15], microrobotics[16], and biological sensing and testing[17]. However, the meniscus-driven movement is usually unidirectional, which is determined by the given wettability and geometry of the floating object. Changing the directionality of the movement requires external stimuli for droplets or particles to overcome the energetic barrier that the meniscus poses, such as adding surfactants[18–20], inducing chemical reactions[21,22], or providing local laser heating[23,24]. In all these approaches, the core concept lies in altering the local interfacial surface tension, consequently changing object wettability or creating local Marangoni flows nearby the object, which change or control the movement of the object. For example, aforementioned meniscus-descending oil droplets on a water bath will quickly climb the meniscus towards the wall once surfactants are added into the water due to the change of interfacial tension[6]. In addition to modifying the surface tension using chemical agents, directly applying laser light onto Janus colloids[16], droplets[24] or liquid marbles[25] can generate Marangoni stresses by inducing asymmetric heating, consequently leading to self-pulsion in a controlled direction. Nonuniform evaporation rates at liquid-air interfaces of volatile menisci or droplets can also lead to a temperature gradient and thereby a thermocapillary interfacial flow, which can reverse the well-known "coffee ring" effect[26]. Above-mentioned approaches have demonstrated great potential for the manipulation of individual objects, but the selective manipulation of a collection of floating objects remains a challenge.

In this work, we present a more feasible approach to manipulate water microdroplets to concurrently ascend and descend a thin-film oil meniscus. We experimentally show that the direction of movement of different-sized droplets can be manipulated using thermo-regulation (*i.e.*, by heating or cooling the substrate). Through numerical simulations that couple heat transfer and fluid flow and confocal fluorescence microscopy experiments, we reveal that a temperature gradient establishes along the oil-air interface that leads to thermal Marangoni convection within the meniscus. A scale analysis based on lateral force balances suggests that the bidirectional droplet movement is caused by the competition of unbalanced capillary forces and shear forces exerted by the convective flow. We finally characterize the influence of meniscus geometry and substrate temperature on the Marangoni convection strength and direction and introduce a modified Marangoni number, which well predicts the magnitude of the flow velocity.

## 2. Expeirmental methods

### 2.1 Sample preparation

To prepare the samples, we first rinsed sapphire windows (1" in diameter with a thickness of 0.02") with acetone, isopropanol, and de-ionized (DI) water in sequence and dried them using compressed nitrogen gas. We chose sapphire as the substrate due to its transparency and relatively high thermal conductivity (40 W/(m·K) as compared to 1.14 W/(m·K) for borosilicate glass), which helps minimize temperature inhomogeneities across the sample during heating and cooling. The cleaned substrates were coated with Glaco Mirror Coat solution (Soft 99 Co., Japan) using spin coating at 600 RPM (Ni-Lo 4 Spin Coater) and then baked on a hotplate at 250 °C for 30 minutes to stabilize the coating layer[27,28]. The processes of coating and heating were repeated twice, resulting in a layer of porous and optically transparent nanostructures (made of hydrophobic nanobeads) with a thickness of ~1 μm. Scanning electron microscope (SEM) images can be found in **Fig. S1** of the **Supplemental Material**. Then, the substrates were impregnated with Krytox GPL 102 fluorinated oil (53 cP at 25°C, Dupont) or silicone oil (19 cP or 48 cP at 25°C, Sigma-Aldrich) at the desired thickness (5 – 50 μm) (see **Supplemental Material, Section 1**). We chose these two oil types due to their prevalence in literature for the fabrication of lubricant-infused surfaces[29–32]. **Table 1** lists the physical properties of the liquids used in this work. We found the densities of the oils not to affect the general results and conclusions presented hereafter. Furthermore, the oils are immiscible with water and tend to spread over water droplets due to positive spreading coefficients, $S_{wo(a)} = \gamma_{wa} - \gamma_{oa} - \gamma_{wo} > 0$, where $\gamma_{wa}$, $\gamma_{oa}$, and $\gamma_{wo}$ denote the interfacial energies between water-air, oil-air and water-oil, respectively[33,34]. However, since the cloaking layer over a water droplet is on the nanoscale[35,36], apparent oil-water-air contact lines can still be visually observed.

*Table 1 Physical properties of liquids at 25 °C*

|   | Liquid | Thermal conductivity $k$ (W/(m·K)) | Surface tension in air $\gamma$ (mN/m) | Viscosity $\mu$ (cP) | Density $\rho$ (g/ml) |
|---|---|---|---|---|---|
| 1 | 20 cSt Silicone oil | 0.15 | 21 | 19 | 0.95 |
| 2 | 50 cSt Silicone oil | 0.15 | 21 | 48 | 0.96 |
| 3 | Krytox GPL 102 | 0.082 | 19 | 54 | 1.86 |
| 4 | Water | 0.609 | 72.8 | 0.89 | 1.0 |
| 5 | Ethylene-glycol | 0.258 | 47.3 | 16.1 | 1.11 |
| 6 | Ethanol | 0.171 | 22 | 1.07 | 0.786 |

### 2.2 Investigation of water microdroplet movement on oil menisci

We first deposited a borosilicate glass sphere (nominal diameter 700-800 μm, Cospheric LLC) at the center of each sample, such that an oil meniscus naturally formed. Then, the meniscus was given at least 20 minutes to equilibrate. At the end of this period, the highest flow velocity within the oil meniscus was less than 2 μm/s (see **Supplemental Material, Section 2**). Subsequently, the substrate temperature was controlled between 2 and 50 °C (± 0.2°C) using a Linkam's PE120 Peltier stage, which has a small hole in its center (diameter: 5 mm) to allow for unobstructed imaging of the specimen. The laboratory environment was at approximately 23 °C. With the goal of observing a wide range of droplet sizes, we used an ultrasonic humidifier (Kelmar, KM-AH026W) containing de-ionized (DI) water to continuously deposit small microdroplets (4 – 28 μm) onto the meniscus. Due to coalescence in the flat oil film region, these sizes naturally evolve to 4 – 290 μm for the experiments on cold substrates (individual experiments lasting 50 seconds or less). To prevent microdroplet evaporation on the substrates at room temperature or higher, we conducted the experiments in an enclosure and increased the local humidity using the humidifier before placing the sample within the enclosure and used both the humidifier and a mist sprayer to deposit droplets (4 – 205 μm). Droplet dynamics were recorded using an sCMOS camera (Pco.edge 4.2 bi) at 15 – 50 frames per second (fps) mounted on an inverted microscope (Nikon Eclipse TE300, with a 10x objective) with diascopic bright-field illumination. The uncertainty in characterizing droplet size and location is ± 1.3 μm.

## 2.3 Fluorescence confocal micro particle image velocimetry (fc-μPIV)

Due to the unavailability of suitable fluorescent dyes for Krytox oils, we used silicone oil (48 cP) dyed with Lumogen Rot 305 (BASF) and labeled with 1-μm Fluoresbrite YG microspheres (Polyscience) at a concentration of 20 μg/g to quantify the meniscus geometry and flow dynamics. The initial oil film thickness was approximately 45 μm. The sample was placed on an inverted scanning confocal microscope (LSM 880 Airyscan, Carl Zeiss, with a 10x, NA = 0.3 objective lens), for which we used the 488 nm and 543 nm laser beams. The substrate temperature was set to 10 °C using the Linkam inverted Peltier stage. We performed horizontal time-series scans at different heights to obtain full flow field data and line z-stack scans to characterize the oil meniscus profile. These experiments were conducted approximately 50 minutes after the glass sphere was deposited.

## 2.4 Characterization Marangoni flow within oil menisci of different sizes and temperatures

The meniscus geometry is contingent on the central object size and the initial oil film thickness. To obtain a greater diversity in oil meniscus geometries, we replaced the central glass sphere with ethylene-glycol droplets (0.25 – 1.2 mm). Compared to the glass sphere, the liquid droplet can deform and squeeze oil from beneath, leading to a faster evolution of the meniscus[37]. The substrates were infused with silicone oil (19 cP) seeded with 1.5 μm polystyrene microparticles at different film thicknesses (6 – 55 μm). We used a digital camera (Canon SL2 with a 65mm, 1-5× macro lens) to determine the meniscus geometry in side-view and monitored the tracer particle movement under an upright microscope (Nikon Eclipse LV100 with a 20x, NA = 0.5 objective lens, equipped with a Photron Mini AX100 high-speed camera). The depth of field is limited at this magnification. Hence, to capture the highest lateral velocity, we focused on and analyzed the particle velocities at different distances from the substrate-oil interface. The maximum measured lateral velocity $U_r$ (projected into the imaging plane) was converted to the interfacial velocity $U_M$ based on the oil meniscus geometry, $U_M = U_r/\cos(\arctan(2h_m/L))$, where $h_m$ is the maximum height of the meniscus (as measured from the top of the flat oil film to the apex of the meniscus) and $L$ is the spanning length of the meniscus (also see **Supplemental Material, Section 3**). The main sources of error for determining the interfacial velocity stem from the measurement of the maximum lateral velocity, $U_r$, of particles and the dimension of the oil meniscus. Since particles move along the sloped oil-air interface, the particles easily get out of focus and blurry towards the end of their trajectories in the top-view observations. In the particle tracking analysis, taking a large interval (~ 1 s) can minimize this error down to 1 μm/s, but might not be able capture the maximum velocity. Here, we chose a large time interval for slow particles and a small interval for rapid particles. The error of characterizing the oil meniscus dimension is less than ± 7 μm. For example, for a small meniscus of $h$ = 280 μm, the uncertainty in the meniscus dimension is ≈ 5 %, which results in the error of $U_M$ of less than 2%. For very small menisci on cooled substrates, where the meniscus might take several days to fully develop, the interfacial Marangoni flow might be cancelled out by the meniscus-feeding flow due to negative Laplace pressure within oil meniscus, leading to a slight underestimation of the (Marangoni) velocity.

## 3. Results and discussion

### 3.1 Size-based bidirectional movement of microdroplets

On a substrate at room temperature, microdroplets of all sizes climb the meniscus (see **Fig. 1(a)**), *i.e.*, they exhibit a unidirectional movement towards the glass sphere. Such meniscus-climbing behavior has been reported previously and is caused by unbalanced capillary forces at the apparent droplet-meniscus-air contact line[8–10]. Surprisingly, however, different droplets simultaneously ascend and descend the meniscus when the substrate is cooled to 8 °C, as shown in **Fig. 1(b)** (also see **Supplemental Material, Video 1**). While relatively larger microdroplets (> 50 µm) still climb the meniscus, smaller microdroplets move in the opposite direction, down the meniscus and away from the central glass sphere. This bidirectional movement contrasts with the expected meniscus-climbing of droplets or small objects.

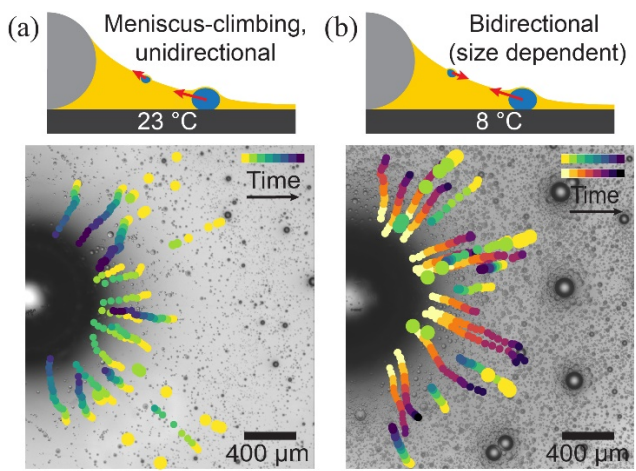

*Figure 1. Microdroplet dynamics on thermally regulated oil menisci. (a) At room temperature, microdroplets climb the oil meniscus that surrounds a 780-µm glass sphere placed on a 27 µm thin oil film of Krytox 102. Tracks of individual droplets (green-hued tracers) are superimposed with a representative still image of the droplet-covered surface. The time interval between two successive tracers is 0.67 s. (b) On a sapphire substrate held at 8 °C, large microdroplets (50 µm – 320 µm) ascend the meniscus (green-hued tracers), whereas smaller droplets descend the oil meniscus (red-hued tracers).*

To rationalize this bidirectional movement on the cooled substrate, we compare different forces acting on the microdroplets. We hypothesize that the larger microdroplets are still attracted towards the central sphere due to the unbalanced capillary forces, whereas the smaller droplets are carried by a flow within the meniscus *via* shear forces. We neglect the meniscus-feeding oil flow arising from the negative Laplace pressure in the curved meniscus[38], since the flow velocity of such capillary suction is nearly an order of magnitude smaller than the observed velocity of the outward flowing droplets, not to mention the opposite flow direction (see **Supplemental Material, Section 2**). Furthermore, the substrate is colder than the upper oil-air interface, ruling out buoyancy-driven flow. Instead, we propose that a temperature gradient (warmer at top, cooler at bottom) establishes along the oil-air interface due to the nonuniform thickness of the curved meniscus and a higher temperature in the surrounding air than on the cooled substrate. Surface tensions of Krytox and silicone oils decrease with increasing temperature, leading to thermocapillary stresses along the oil-air interface. The resulting microscopic thermal Marangoni convection pulls oil from the apex of the meniscus to the flat film region. We thus expect that two mechanisms play a competitive role in determining size-dependent droplet movement: unbalanced capillary forces and thermocapillary convection. Although thermal Marangoni convection has been considerably studied in volatile liquid drops and menisci[26,39–43], thermocapillary dynamics in a nonvolatile meniscus have received much less attention.

### 3.2 Simulation of heat transfer and fluid dynamics within an oil meniscus

To validate our hypothesis, we examined the oil-air interfacial temperature profile of the meniscus surrounding a glass microsphere on a cold substrate (10 °C). Due to experimental limitations of measuring the microscopic meniscus temperature with probe-based sensors or using infrared (IR) imaging, we instead performed a conjugate heat transfer simulation of the solid sphere and its oil meniscus *via* COMSOL Multiphysics, where we considered the coupling of non-isothermal flow and Marangoni effects (for more simulation details, see **Supplemental Material, Section 4**). The meniscus geometry was imported from confocal microscopy measurements. **Figure 2(a)** shows the distributions of the interfacial temperature $T_i$ and the lateral velocity $U_r$ along the oil-air interface ($r >$ 390 μm, where $r =$ 0 is the contact point of the sphere with the substrate). The highest interfacial temperature exists close to the apex of the meniscus and decreases towards the flat film region. The strongest temperature gradient at $r \approx$ 650 μm also results in the highest lateral velocity of $\approx$ 15 μm/s. At larger distances from the central sphere, the velocity gradually decreases to zero due to the diminishing interfacial temperature gradient. Generally, the interfacial velocity decreases with an increase in thermal conductivity of the central object (see **Supplemental Material, Section 5**).

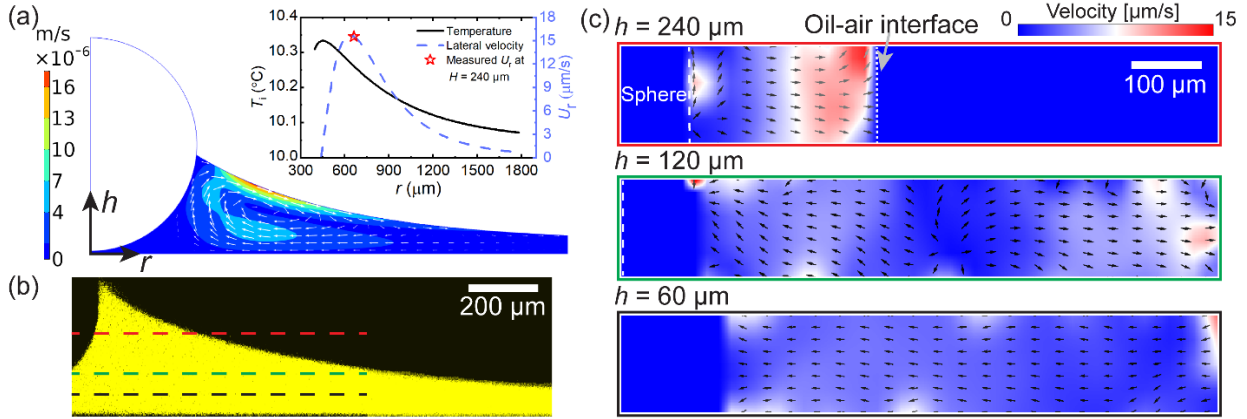

*Figure 2. Visualization of flow pattern inside an oil meniscus. (a) Simulation of heat transfer and fluid dynamics within a silicone oil (48 cP, 45 μm flat film thickness) meniscus surrounding a glass sphere with a substrate temperature of 10°C. The insert shows the oil-air interfacial temperature and lateral velocity from the meniscus apex to the flat film region. (b) Side-view confocal fluorescence image of a typical silicone oil meniscus (yellow) surrounding the glass sphere. (c) Horizontal velocity fields measured by fluorescence confocal micro particle image velocimetry at three different heights from the oil-substrate interface, as marked in (b).*

### 3.3 Experimental flow field measurements

To experimentally confirm the flow dynamics within the oil meniscus, we dyed and labeled silicone oil and used fluorescence confocal micro particle image velocimetry (fc-µPIV) to trace the 1 µm fluorescent microspheres. **Figure 2(b)** shows a typical cross-sectional geometry of the meniscus. The three dashed lines represent the heights at which we performed horizontal time-series scans to obtain flow field data, namely at 60 µm, 120 µm, and 240 µm away from the substrate-oil interface, respectively. As shown in **Fig. 2(c)**, at $h = 240$ µm, we observe a radially outward flow, with the velocity increasing towards the oil-air interface. The experimentally measured lateral velocity at the oil-air interface agrees well with the simulation (red star in the insert of **Fig. 2(a)**). At $h = 120$ µm, the flow is outward near the oil-air interface, but reverses its direction nearby the microsphere, indicating a strong circulating flow in the vertical direction. To compensate for the Marangoni-induced outward flow near the interface, a uniform inward flow establishes close to the substrate at $h = 60$ µm, *i.e.*, a convective vortex structure establishes. Once the cooling is turned off and the substrate returns to room temperature, the convective flow gradually disappears. When heating the substrate, we observe a similar convection roll, but in the reversed direction (see **Supplemental Material, Section 6**), which further substantiates our explanation of thermal Marangoni flow. In the following, we call the convection roll with surface flow from the microsphere to the flat film region as positive and the reversed direction as negative. Positive and negative convection rolls are expected to play distinctive roles by fueling or reversing meniscus-climbing microdroplet locomotion.

### 3.4 Geometries of water microdroplets at the oil-air interface

Now, we return to our original question: why do droplets of different sizes move in opposite directions on a cooled substrate? Let us first examine the geometries of these microdroplets interacting with an oil meniscus, which heavily depend on the relationship between the droplet diameter $d$ and the local oil film thickness $h$. As illustrated in **Fig. 3(a)**, we can categorize two kinds of droplets: droplets with $d > h$ and $d < h$, respectively. All droplets, irrespective of their size, locally deform the oil meniscus, establishing their own small menisci. In the following, we call this small droplet-centric oil meniscus "secondary meniscus" (yellow solid lines in **Fig. 3(a)**) and the larger meniscus of the central object "primary meniscus" (yellow dashed line). On Krytox oil (high density), droplets with $d < h$ will float at the oil-air interface[44]. Even on silicone oil, which has a slightly lower density than water, small droplets will initially float at the oil-air interface and then slowly sink into the bulk. The timescale of sinking is at least one order of longer than the timescale of the convective flow[45], allowing us to consider these microdroplets to be floating as well. For both types of oil, these floating microdroplets cause only a minor disturbance to the interface, as shown **Fig. 3(b)**, and maintain an approximately spherical shape. The outer red dashed circle in the photograph indicates the droplet profile underneath the oil-air interface (*i.e.*, within the primary oil meniscus) and the inner white dotted circle highlights the apparent water-oil-air contact line, featuring a small contact line radius, $r_m < d/2$. We call it "apparent" contact line, since technically due to the cloaking nature of the oils there is no real three-phase contact line, but nonetheless the inflection in the surface profile is pronounced enough to cause similar effects to a true contact line (such as capillary forces, *etc.*). On the other hand, droplets with $d > h$ are supported by the underlying solid substrate and hence protrude the oil-air interface to a greater extent, leading to a significant secondary meniscus, as shown in **Fig. 3(b)**, for which the apparent contact line dimension becomes comparable to that of the droplet (*i.e.*, $r_m \sim d/2$).

As the droplet moves towards the central object, $r_m$ will continuously decrease as the droplet gradually immerses into the meniscus due to substrate-facing capillary forces and eventually floats just underneath the oil-air interface.

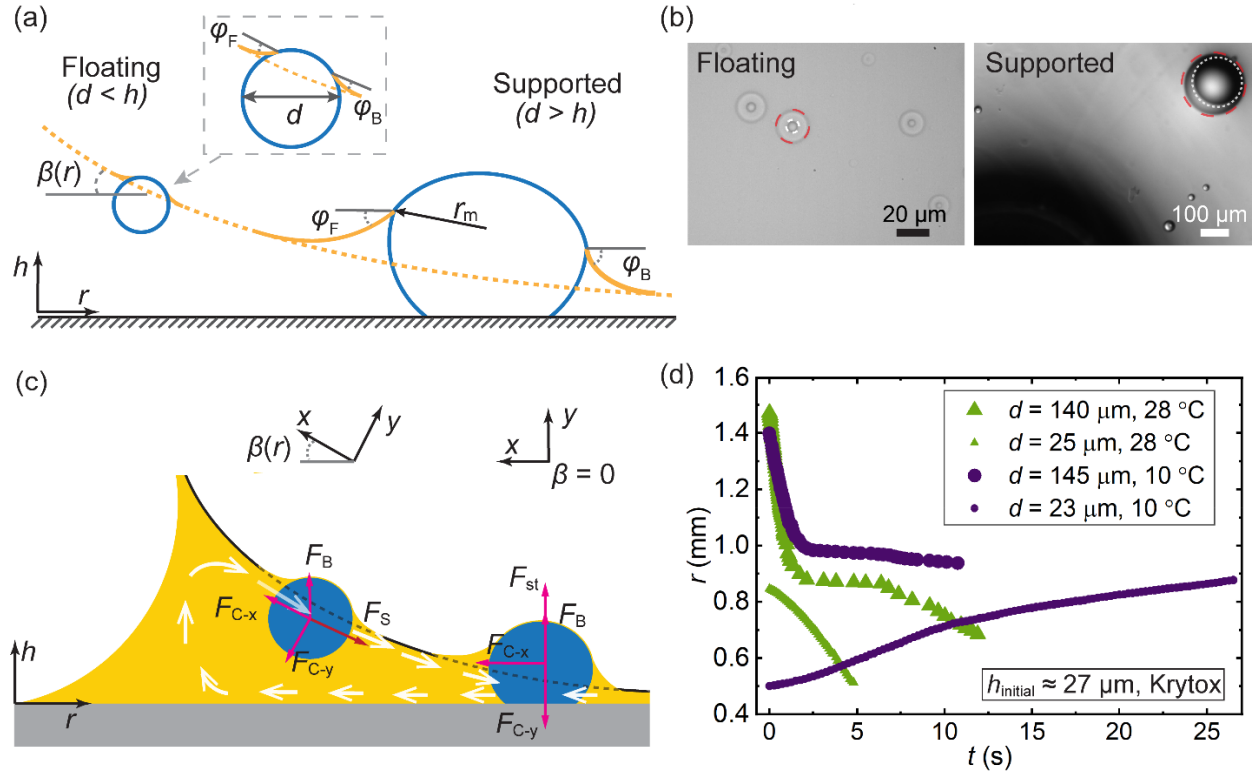

Figure 3. Competition of unbalanced capillary and shear forces dictates microdroplet movement. (a) Schematic showing the geometries of floating (d < h) and supported droplets (d > h) within an oil meniscus (yellow dashed line). (b) Top-view photographs of small microdroplets (d < 20 μm) floating at the oil-air interface of a 60 μm Krytox 102 oil film and a partially immersed larger microdroplet (d ≈ 210 μm) supported by the substrate. White dotted circles indicate the apparent water-oil-air contact lines and red dashed lines represent the droplet profiles submerged in the oil. The large droplet is moving within the primary oil meniscus, causing the non-symmetric appearance. (c) Schematic showing the force balance in a droplet-centric x – y coordinate system, where unbalanced capillary forces due to the overlap of primary and secondary menisci compete with shear forces from the convective oil flow on a cooled substrate. (d) Representative trajectories of large and small microdroplets on cooled (10 °C) and heated (28 °C) substrates, respectively.

### 3.5 Force analysis for droplets on a stationary meniscus

To better understand how these two kinds of droplet and meniscus morphologies influence the droplet dynamics, we perform force analyses for the supported and floating droplets, respectively, under isothermal conditions, *i.e.*, in the absence of flow within the meniscus. A droplet-centric $x - y$ coordinate system is established by rotating the $h - r$ coordinate by the angle $\beta$ between the direction of droplet movement and the horizontal (for floating droplets, $\beta$ is approximately the local slope of the undisturbed primary meniscus), as shown in **Fig. 3(c)**. The overlap of the primary and secondary menisci causes a well-known attractive capillary force $F_{C-x}$ on the droplet towards the central solid sphere[8,10,46,47]:

$$F_{C-x} \approx 2\pi r_m \gamma_{oa}(\cos\varphi_F - \cos\varphi_B) = F_C \cdot (\cos\varphi_F - \cos\varphi_B), \quad (1)$$

where $\varphi_F$ and $\varphi_B$ are defined as the angles between the tangent to the secondary (droplet-based) meniscus and the $x$-axis (see **Fig. 3(a)**).

Due to the density mismatch, $\Delta\rho$, between the droplet and the oil, there exists a buoyancy force, $F_B \sim d^3 \Delta\rho g$ that acts on the droplet. Due to the complex geometry of large, substrate-supported droplets and their menisci, it is impossible to determine $F_B$ exactly, but that does no harm to the general analysis in the subsequent sections. For large droplets protruding oil film, the support force from the substrate can be denoted as $F_{st}$. Balancing the forces acting on the droplet in the $x - y$ coordinate system yields:

$x$ – axis:
$$F_{driving} = F_C \cdot (\cos\varphi_F - \cos\varphi_B) + F_B \cdot \sin\beta, \quad (2)$$

$y$ – axis:
$$F_C \cdot (\sin\varphi_F + \sin\varphi_B)/2 - F_B \cdot \cos\beta + F_{st} = 0, \quad (3)$$

Large droplets move primarily laterally towards the central object, so that $\beta \approx 0$ and the droplet movement is only driven by the unbalanced capillary force (there is a slight, but negligible, upward motion to compensate for the change in buoyancy forces, until the droplet barely touches the substrate and transitions into becoming a "small floating" droplet – more see below. Also note that the notion of a "large" and "small" droplet is always in relation to the local thickness of the primary meniscus, and not an absolute value). The "secondary meniscus" geometry of protruding droplets varies with the lubricant availability, which is determined by the primary meniscus curvature. From experimental observations, we estimate that the angle difference $(\cos\varphi_F - \cos\varphi_B) \sim 10^{-1}$, propelling the droplet towards the central sphere (see **Supplemental Material, Section 7**). From $\frac{F_B}{F_C} = Bo \sim \frac{|\Delta\rho|gd^3}{\gamma_{oa}r_m} \sim 10^{-2}$, where $d \sim r_m \sim 10^{-4}$, and the simultaneous support from the substrate, it becomes apparent that these large droplets will deform significantly due to the downward-facing capillary forces[35].

As the droplet moves towards the central sphere, these downward facing capillary forces lead to the gradual immersion of the droplet into the primary oil meniscus. During this process, the droplet relaxes its shape until it becomes nearly spherical when its diameter is approximately the same as the local meniscus height, i.e., $d \approx h$. At this point, the support force $F_{st}$ disappears. As the droplet moves into the meniscus, the size of the apparent droplet-oil-air contact line also decreases. For sufficiently small $r_m$, the primary meniscus can locally be approximated as a planar interface, resulting in the angle difference between $\varphi_F$ and $\varphi_B$ to gradually decrease. On the other hand, the slope of the primary meniscus, i.e., $\beta$, increases. For such a droplet ($d \approx h \sim 10^{-5}$ m) to satisfy eq. (3), it requires $(\sin\varphi_F + \sin\varphi_B)/\cos\beta \sim 10^{-4}$, where from the shape of an undisturbed primary meniscus we know that $\cos\beta \sim 10^{-1}$. This means that the angles $\varphi_F$ and $\varphi_B$ are extremely small, as expected. The angle difference $(\cos\varphi_F - \cos\varphi_B)$ then mainly stems from the curvature of the primary meniscus, so it can be approximated as $(\cos\varphi_F - \cos\varphi_B) \sim (\cos\beta_F - \cos\beta_B)$ (see **Supplemental Material, Section 7**).

For even smaller droplets, that are fully or partially submerged, as discussed in **section 3.4**, $Bo < 10^{-3}$, $(\cos\varphi_F - \cos\varphi_B) \sim 10^{-3}$, and $\sin\beta \sim 10^{-1} - 10^0$. Hence, unbalanced capillary forces continue to dominate over the $x$-component of the buoyancy force, meaning that the oil density or orientation of the sample do not influence droplet dynamics. This was confirmed experimentally for an inverted setup using Krytox oil, in which floating droplets still ascended the oil meniscus, i.e., moved with gravity, instead of descending the meniscus (against gravity), as would be the case for a buoyancy-driven phenomenon due to the higher density of Krytox as compared to water.

### 3.6 Force analysis for droplets on a meniscus subject to Marangoni convection

Having established that capillary forces dominate for both supported and floating droplets on a stationary meniscus, we can now introduce meniscus convection and determine its effect on droplet dynamics. On a heated or cooled substrate, the capillary forces discussed in the previous section compete with forces caused by the oil flow, specifically shear on the droplet. The Reynolds number for the Marangoni convection $Re = \rho U_M h_m/\mu_o \approx 10^{-4} - 10^{-3}$, where $\rho$ is the oil density, $U_M$ is the characteristic velocity of the flow ($O(10 - 100$ μm/s)), $h_m$ is the height of the meniscus at its apex ($O(100$ μm)), and $\mu_o$ is the oil viscosity, meaning the flow can be approximated as Stokes flow with negligible inertia. The characteristic velocity of Marangoni flow at the oil-air interface scales as $U_M \sim \gamma_T \Delta T/\mu_o$, where $\gamma_T = d\gamma_{oa}/dT$ represents the temperature-dependence of the oil surface tension and $\Delta T$ is the temperature difference along the meniscus[40]. For the sake of simplicity, we consider a uniform flow surrounding the small floating droplets near the interface. The shear force exerted by the surrounding fluid on a floating microdroplet ($d \ll h$) then becomes

$$F_s \sim 3\pi\mu_o d U_M \sim 3\pi d \gamma_T \Delta T, \tag{4}$$

which is in the same direction as $F_{C-x}$ on a heated substrate, but opposes $F_{C-x}$ on a cooled substrate (see **Fig. 3(c)**). The direction of movement of these water microdroplets depends thus on a competition of $F_{C-x}$ and $F_s$, so we define a dimensionless number

$$X = \frac{F_s}{F_{C-x}} \sim \frac{\gamma_T \Delta T}{\gamma_{oa}} \cdot \frac{d}{r_m} \cdot \frac{1}{\cos\varphi_F - \cos\varphi_B}, \tag{5}$$

which includes a forcing term (~ substrate temperature) and a geometric term (~ droplet size). For $X > 1$, we would expect the droplet to follow the flow, whereas for $X < 1$, the droplet would climb the meniscus, irrespective of the flow direction. The temperature-dependence of the oil surface tension $\gamma_T \sim -5\times10^{-5}$ N/(m·K) [48], so the first term, $\gamma_T \Delta T/\gamma_{oa}$, is on the order of ~ $10^{-3}$ for a typical temperature difference of 0.5 K along the oil-air interface. As discussed above, for floating droplets, $r_m < d$ and $(\cos\varphi_F - \cos\varphi_B) \sim 10^{-3}$. Consequently, the dynamics of these small floating droplets are dominated by shear forces ($X > 1$) and they descend the meniscus for positive convection rolls (cooled substrate). For larger droplets that protrude the oil film, $r_m \sim d$ and $(\cos\varphi_F - \cos\varphi_B) \sim 10^{-1}$. Furthermore, the size of the larger droplets exceeds the local height of the oil meniscus ($d > h$), meaning the droplet experiences shear forces from the convection roll in opposite directions at the upper and lower part of the droplets, which approximately cancel out (the vortex might lead to droplet rotation; however, this was not observed experimentally). Consequently, large droplets display consistent meniscus-climbing due to dominating capillary forces ($X < 0.01$). To illustrate the influence that the competition of lateral capillary and shear forces has on droplet motion, we selected four representative droplets – two small floating droplets (≈ 24 μm) and two large droplets (≈ 140 μm) on two samples held at 10°C and 28°C, respectively – and plotted their trajectories over time. **Figure 3(d)** confirms that the direction of movement of small droplets (smaller symbols) follows the directional change of the convection roll at the different temperatures. They ascend the oil meniscus at 28°C, *i.e.*, the distance $r$ to the central sphere decreases with increasing time, but descend the meniscus at 10°C (increasing $r$). Larger droplets, however, are nearly unaffected by the Marangoni convection and at early times (when $d > h$ and $r_m \sim d$) climb the meniscus at both temperatures with velocities one order of magnitude higher than those of the small droplets. As the large droplets move closer to the central object, the unbalanced capillary forces gradually abate due to a decreasing $r_m$ together with a decrease in (cos $\varphi_F$–cos $\varphi_B$), at which point shear forces from the convective meniscus flow become increasingly important. The large droplet at 28°C continues to ascend the meniscus at a velocity similar to that of the small droplet due to the shear stress, however, the one at 10°C becomes stationary due to the cancellation of forces. We expect that droplets will switch their moving direction when $X \sim 1$, which is determined by the interplay of imposed temperature gradient, meniscus profile, droplet size, and the distance between central sphere and moving droplet. For the experiments of **Figs. 1(b)** and **3(d)**, the transition radii of water droplets are on the order of $10^{-5}$ m. It is worthy to note that a larger droplet ($d > h$) will always ascend the primary meniscus, irrespective of its initial radial distance from the central object (see **Supplemental Material, Section 8**).

### 3.7 Prediction of Marangoni flow within oil meniscus of different geometries and temperatures

We envision this thermally activated bidirectional movement to enable the manipulation of droplets and potentially small particles on thin liquid interfaces for applications in self-assembly, microrobotics, and bio-medical microfluidics. To provide better guidelines on the influence of meniscus geometry and substrate temperature, we conducted a series of additional experiments. To account for both the height and the spanning length of the meniscus, we define $\sqrt{h_m^2 + L^2}$ as the characteristic length of the oil meniscus. **Figure 4(a)** shows a regime map of the interfacial (Marangoni) velocity relative to the substrate temperature and indicates the direction of the Marangoni convection. As expected, no flow is observed for an isothermal setup. For very small menisci, irrespective of the substrate temperature, convection is also absent due to negligible temperature gradients (for example, a long but shallow meniscus has roughly the same thermal resistance and hence oil-air interfacial temperature). Convective flow patterns, whose directions are determined by the substrate temperature $T_s$ relative to the environmental temperature $T_{env}$, start to emerge at $\sqrt{h_m^2 + L^2} \approx 0.4$ mm (see **Supplemental Material, Video 2**). Higher temperature differences and larger menisci contribute to stronger convective flow. Due to the inverse temperature dependence of the oil viscosity, the velocity magnitude is lower for positive (cold substrate) than negative (hot substrate) convection rolls at otherwise same conditions. It is also interesting to note that negative convection rolls also appear within oil menisci surrounding millimetric volatile droplets, such as water and ethanol, even on substrates held at room temperature, due to evaporative cooling near the oil-volatile droplet-air contact line (see **Supplemental Material, Video 3**). Such convection within the oil meniscus has been largely ignored in past studies on the evaporation of (suspension) droplets on oil-coated surfaces[38,49–52], but we propose that the oil convection can affect droplet (evaporation) dynamics and final particle deposition. This mechanism also allows for a fully passive transport of small objects on menisci and could potentially be used for complex self-assembly of two-species particles (dispersed in droplet and oil phases).

To predict the magnitude of the surface flow velocity of the convection cells, we introduce a modified Marangoni number based on the prescribed temperature difference between the substrate and the room environment:

$$Ma = -f \frac{\gamma_T (T_s - T_{env})(h_m^2 + L^2)^{1/2}}{\mu_o \alpha}, \tag{6}$$

where $f$ is a fitting factor to account for the mismatch in using the environmental temperature as opposed to the actual oil interfacial temperature and $\alpha \approx 5 \times 10^{-8}$ m$^2$/s is the thermal diffusivity of the oil. It is important to note here that the *Ma*-number from **eq. (6)** is defined with the $\Delta T$ of the meniscus, not that across an individual droplet. This is in contrast to, for example, millimetric droplets on a temperature-gradient surface which self-propel due to thermocapillary forces across the droplet[53,54]. Here, we are interested in determining the magnitude of the convective flow within the oil meniscus, as this carries the floating microdroplets up or down the meniscus, as discussed in **section 3.6**. We calculated the *Ma* number for each data point of **Fig. 4(a)**, using $f = 0.175$ based on findings from the numerical simulations (see **Supplemental Material, Section 9**). The absolute values of the interfacial velocity *vs. Ma* numbers are plotted in **Fig. 4(b)**. All data points collapse onto one line scaling as $U_M \sim Ma$, confirming that the interplay of temperature gradient and meniscus size determines the strength of the Marangoni convection. **Equation (6)** thus provides a good prediction of the Marangoni flow strength.

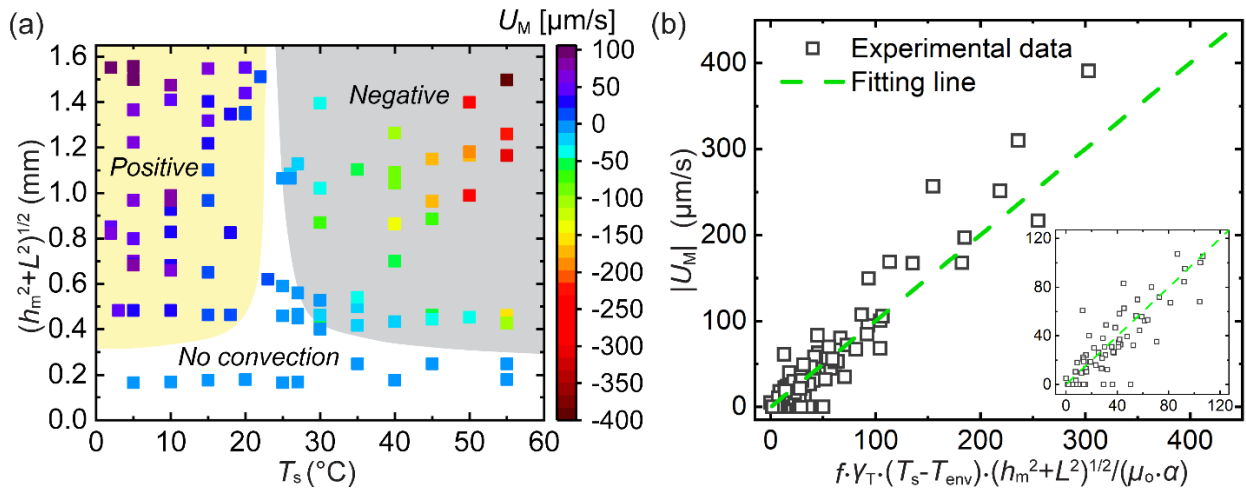

*Figure 4. Dependence of Marangoni flow on substrate temperature and oil meniscus geometry. (a) Regime map of the interfacial flow velocity relative to the substrate temperature. The experiments were conducted in an open room environment at 23 °C. (b) The magnitude of the interfacial Marangoni flow shown in (a) is directly proportional to the proposed Marangoni number (eq. (6), where $f = 0.175$).*

### 3.8 Influence of bidirectional droplet propulsion on dropwise condensation heat transfer

As seen in **Fig. 4(b),** a very small Marangoni number and hence a very small temperature difference between substrate and environment is sufficient to cause convection within the meniscus. Consequently, Marangoni convection within the oil and thereby a bidirectional movement of droplets (and possibly particles) can easily occur in many applications that experience a vertical temperature gradient. A good example is water dropwise condensation on lubricant-infused surfaces (LISs), where the temperature of the cold substrate is lower than that of the hot vapor in order to initiate phase change. To illustrate the importance of bidirectional droplet motion during dropwise condensation, we conducted a condensation test in which we periodically supplied water vapor ($\approx 40$°C; using nitrogen as carrier gas) to a cold lubricant-infused surface ($\approx 10$°C). In-between experimental runs, water was allowed to largely evaporate. In **Video 4** from the **Supplemental Material**, a $\approx 600$-μm water droplet was left on the surface from previous condensation and coalescence events. At this relatively high vapor-substrate temperature difference (≈30°C), nucleation occurs not only in the thin film region in-between larger droplets, so-called oil-poor regions[11], but also on the oil menisci themselves. The outward interfacial Marangoni flow can radially distribute these meniscus-nucleated droplets to regions further away from the central droplet, *i.e.*, towards oil-poor regions. Once the droplets arrive at the oil-poor regions, they will rapidly grow and coalesce with neighboring droplets due to a combination of reduced thermal and coalescence resistance (there is less oil that needs to be drained prior to water droplet coalescence). This unique dispersing mechanism of nucleated droplets has never been reported before and could enhance our understanding of dropwise condensation on LISs.

## 4. Conclusions and outlooks

In summary, we introduced a novel bidirectional movement (ascending and descending) of microdroplets on an oil meniscus when the underlying substrate was cooled below room temperature. The interplay of a vertical temperature gradient and the curved meniscus establishes a temperature gradient along the oil-air interface and initiates Marangoni convection within the meniscus. For small droplets, the shear forces exerted by the flow overcome unbalanced capillary forces, leading to the transport of floating microdroplets down the meniscus. However, larger microdroplets still climb the meniscus due to stronger unbalanced capillary forces originating from overlapping menisci and nullified Marangoni convection effects. We showed that the convection direction and magnitude can be tuned by varying the substrate temperature and the meniscus profile. Our results emphasize the significance of Marangoni stresses at the interface even of nonvolatile menisci and identify their influences on manipulating microscale droplets, particles or even microorganism on curved interfaces, which potentially opens a new pathway for enhancing condensation heat transfer, sorting particles, or enabling biological sensing and testing.

## 5. Acknowledgement

This work is supported by the National Science Foundation under Grant No. 1856722. We thank David Quéré for insightful discussions.

# Supplemental Material

## Marangoni-Induced Reversal of Meniscus-Climbing Microdroplets

Jianxing Sun[1], Patricia B. Weisensee[1,2,*]

[1]Department of Mechanical Engineering & Materials Science, Washington University in St. Louis, St. Louis, Missouri 63130, USA

[2]Institute of Materials Science and Engineering, Washington University in St. Louis, St. Louis, Missouri 63130, USA

*Corresponding author: p.weisensee@wustl.edu

## Content

S1. Characterization of Glaco and oil coatings

S2. Capillary-suction-induced oil meniscus flow velocity

S3. Conversion of radial velocity $U_r$ into interfacial velocity $U_M$

S4. Simulation of heat transfer and fluid dynamics within an oil meniscus

S5. Influence of thermal conductivity of the central sphere on oil convection

S6. Negative convection rolls in an oil meniscus on a heated substrate

S7. Estimation of (cos $\varphi_F$ – cos $\varphi_B$) for small floating droplets

S8. Dynamics of droplets initially dispensed closer to the meniscus top

S9. Determination of fitting factor $f$

**Video 1** Comparison of microdroplet movements on an oil meniscus (Krytox GPL 102, 53 cP) surrounding a 780-µm borosilicate glass sphere on room temperature and cooled (8 °C) sapphire substrates, respectively.

**Video 2** Marangoni convection in a silicone oil (19 cP) meniscus surrounding a non-volatile ethylene-glycol droplet ($D ≈ 700$ µm) on heated (50 °C) and cooled (5 °C) sapphire substrates (both at steady state).

**Video 3** Marangoni convection in an oil meniscus surrounding a water droplet sitting on a glass bottom petri dish pre-wetted with 19 cP silicone oil at room temperature. The dish cover is initially closed and then opened, which changes the water droplet's evaporation rate and latent cooling.

**Video 4** Bidirectional movements of condensed microdroplets during water condensation on a cooled sapphire substrate infused with 9.3 cP silicone oil. The temperatures of substrate and water vapor (using nitrogen as carrier gas) are set at 10 °C and 40 °C, respectively.

## S1. Characterization of Glaco and oil coatings

The scanning electron microscope (SEM) images of **Fig. S1(a)** show a layer of the uniform porous Glaco coating in cross-sectional and top views. The contact angle for a millimetric water droplet on the Glaco-coated surface is 165 ± 3°. We applied thin layers of oil (5 – 27 μm) on the textured surfaces *via* spin coating at 600-1000 rpm. To coat thicker oil films (28 – 50 μm), we applied a known amount of oil to the sample and let the oil even out overnight (oil has a positive spreading coefficient on Glaco and hence fully wets the sample). We weighed the mass of the oil using an analytical balance (OHAUS Pioneer PA64) and calculated the oil film thickness based on the substrate dimensions. **Figure S1(b)** shows the correlation between spin speed, oil mass, and oil film thickness. The accuracy of the correlation and the repeatability of the oil thickness using both approaches were confirmed using a confocal fluorescence microscope (LSM 880 Airyscan, Carl Zeiss), where silicone oils were dyed with Lumogen Rot F305. Based on the similarity in viscosity and surface tension, we assume that the thickness of the Krytox oil is approximately the same.

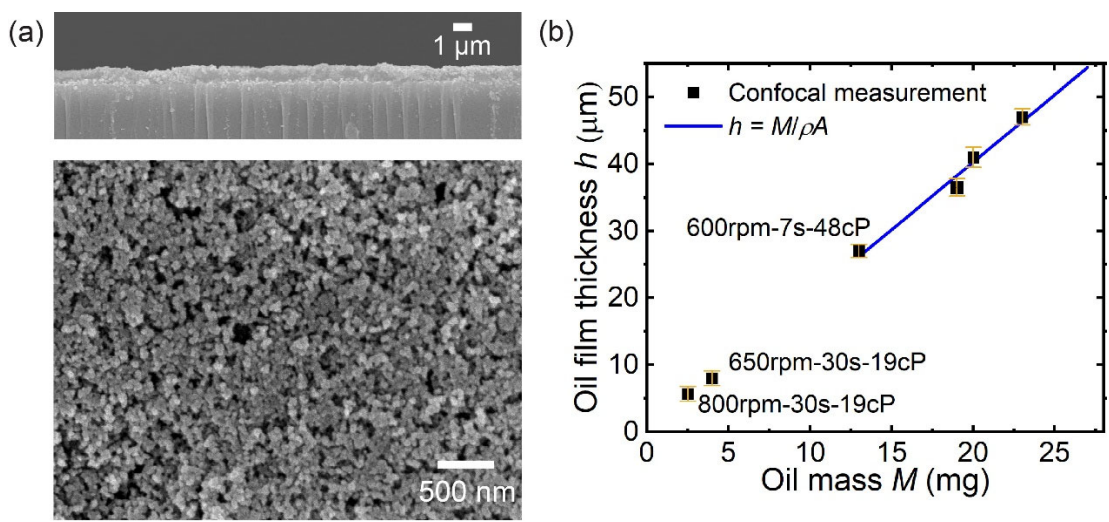

*Figure S1. Characterization of the superhydrophobic coating and oil film thickness. (a) Scanning electron microscope (SEM) images of the Glaco coating in cross-sectional (top) and top (bottom) views. (b) Oil film thickness vs. mass of oil. The oil film thickness was measured using confocal microscopy to validate the accuracy and repeatability of the coating methods.*

## S2 Capillary-suction-induced oil meniscus flow velocity

The dynamic growth and final size of the oil meniscus significantly depends on the amount of lubricant available in the system[1,2]. In our experiments, the initial oil film thickness of the samples is thick and central objects are sub-millimetric, featuring an oversaturated system. The growth speed of the meniscus is thus relatively fast. As shown in **Fig. S2**, the pressure-driven flow velocity quickly decreases to 2 μm/s at $t$ = 20 mins, where $t$ = 0 min represents the moment when the glass sphere is placed on the substrate.

- 2 -

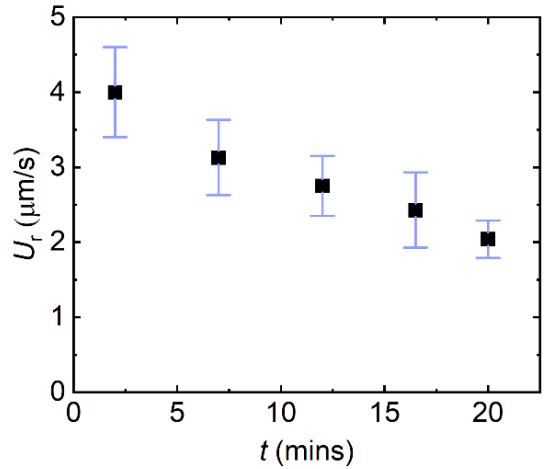

*Figure S2. Capillary-suction-induced oil meniscus flow velocity changes over time. The velocity was measured via confocal microscopy, where the silicone oil (19 cP) was dyed with Lumogen Rot F305 and labeled with a low concentration of 1-μm Fluoresbrite YG microspheres. The initial oil film thickness was about 41 μm. The central object was a 750-um glass sphere and the substrate was kept at room temperature. At minimum of five locations were measured for each time point and the error bar represents the velocity range.*

### S3 Conversion of radial velocity $U_r$ into interfacial velocity $U_M$

We experimentally measured the radial velocity $U_r$ of particles inside oil menisci and found that the maximum values of $U_r$ primarily occurred in the green shade region marked in **Fig. S3(a)**. The interfacial velocity $U_M = U_r / \cos\alpha$, where $\alpha$ denotes the meniscus slope angle. Based on geometrical considerations, we can approximate $\beta \approx \psi = \arctan(2h_m/L)$. **Figure S3(b)** shows a meniscus profile in side-view and the location of maximum $U_r$, substantiating the geometry relationship.

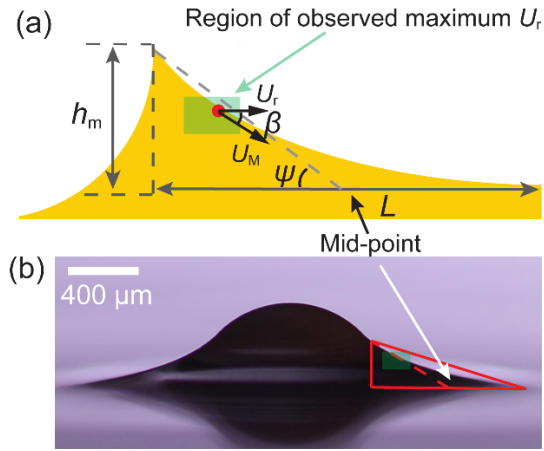

*Figure S3. Conversion of $U_r$ into $U_M$ based on oil meniscus geometry. (a) Schematic of location of maximum $U_r$ within oil meniscus. (b) An experimental image from side-view shows the profiles of a central droplet (Glycol) and a 19-cP silicone oil meniscus.*

## S4 Simulation of heat transfer and fluid dynamics within an oil meniscus

To interpret the fluid flow and temperature field within the oil meniscus, we simulated the conjugate heat transfer and fluid dynamics using COMSOL Multiphysics. As shown in **Fig. S4(a)**, we used a 2D axisymmetric geometry that includes the glass sphere and its oil meniscus. The meniscus profile was imported from the experimental measurements using confocal microscopy. In the simulation, we assigned a no slip boundary condition to the sphere-oil and substrate-oil interfaces and defined the oil-air boundary as a free interface. The Marangoni effect is included by assigning a boundary condition at the oil-air interface[3]:

$$\vec{n} \cdot [-pI + \mu_o(\nabla \boldsymbol{u} + (\nabla \boldsymbol{u})^T)] \cdot \vec{t} = -\vec{t} \cdot \gamma_T \nabla T, \tag{S1}$$

where $\vec{n}$ is the unit outward normal to the surface, $p$ is the pressure, $\mu_o$ is the oil viscosity, $\boldsymbol{u}$ is the velocity vector, $\vec{t}$ is the orthonormal tangent vector to the interface, $T$ is the temperature, and $\gamma_T$ is the temperature-dependent surface tension ($\approx -5.8 \times 10^{-5}$ N/(m·K)) of the silicone oil[4]. The term $[-pI + \mu_o(\nabla \boldsymbol{u} + (\nabla \boldsymbol{u})^T)]$ represents the stress tensor.

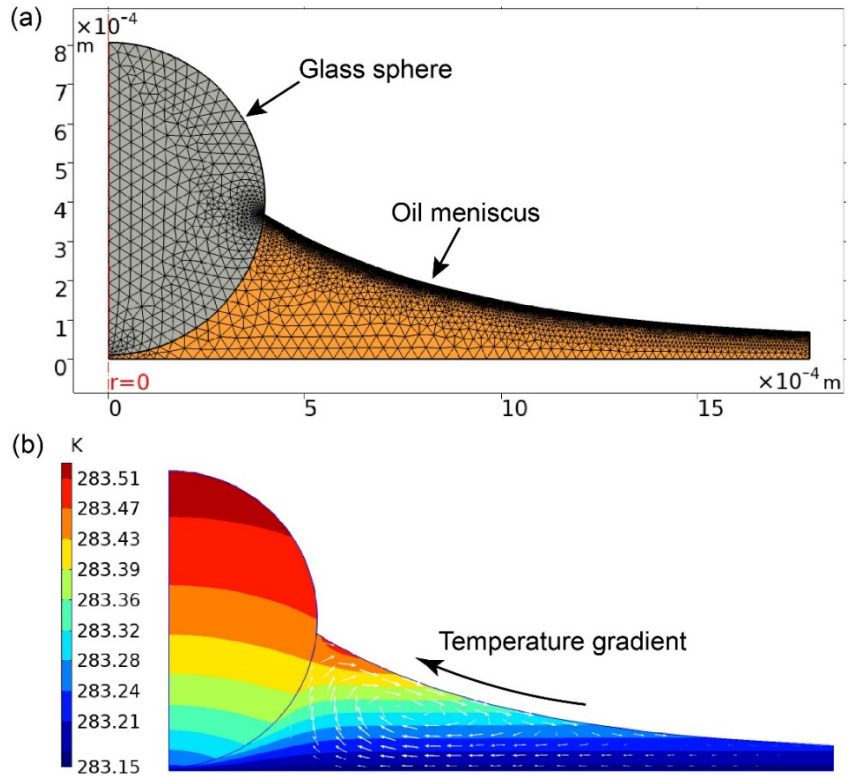

*Figure S4. Simulation domain and results. (a) Mesh of the system under consideration, including two domains: solid sphere and liquid oil meniscus. (b) Temperature distribution within the sphere and oil meniscus in cross sectional view. The white arrows are velocity vectors.*

The temperature boundary condition for the substrate-oil interface was set as constant temperature (283.15 K) in accordance with the experiments. At the sphere-air and oil-air interfaces, a forced convection thermal boundary condition with the room at 296.15 K (23 °C) was imposed, for which the indoor air flow velocity was assumed to be 0.1 m/s. The silicone oil has a very low vapor pressure (<5 mm Hg at 25 °C), so its evaporation was neglected. The convective flow within the meniscus and the oil-air interfacial temperature distribution were obtained by solving the energy conservation equation in the solid sphere and the mass, momentum, and energy conservation equations in the oil meniscus:

$$\nabla \cdot (\rho \boldsymbol{u}) = 0, \tag{S2}$$

$$\nabla p = \mu_\text{o} \nabla^2 \boldsymbol{u}, \tag{S3}$$

$$\rho c_\text{p} \boldsymbol{u} \cdot \nabla T - \nabla \cdot (k \nabla T) = 0, \tag{S4}$$

where $\rho$ is the density, $\nabla p$ is the pressure gradient, $c_\text{p}$ is the specific heat capacity at constant pressure, and $k$ is the thermal conductivity. **Figure S4(b)** shows that a vertical temperature gradient establishes along the oil-air interface, inducing a Marangoni convection roll.

## S5 Influence of thermal conductivity of the central sphere on oil convection

We experimentally demonstrated that the flow field within the oil menisci depends on the oil meniscus geometry and the substrate temperature. To examine the role of the thermal conductivity of the central sphere, we varied the thermal conductivity of the central sphere and used representative values of common materials: 0.3 W/(m·K) (Teflon), 0.6 W/(m·K) (water), 1.14 W/(m·K) (borosilicate glass) and 15 W/(m·K) (stainless steel), while all other parameters stayed the same. **Figure S5** shows that the interfacial velocity decreases with an increase in thermal conductivity. When $k > 0.6$, the interfacial flow direction is first inward and then becomes outward, featuring a stagnation point in the region of the highest temperature.

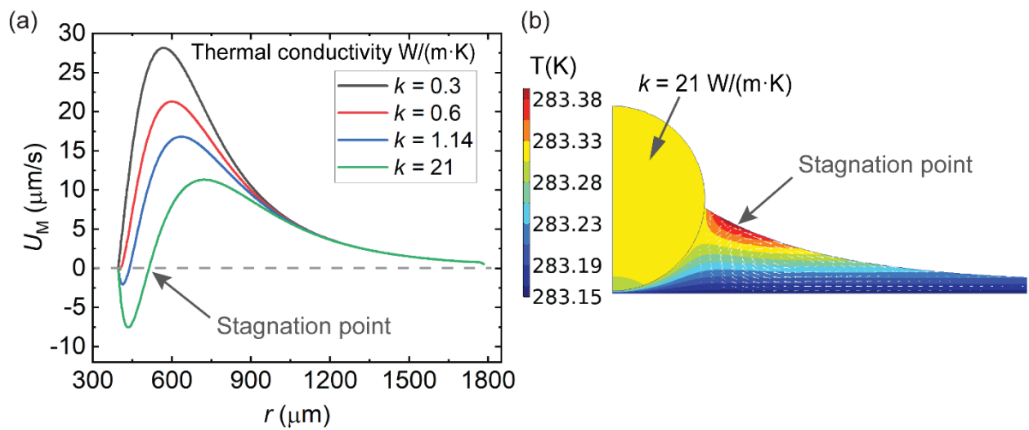

*Figure S5. Simulation of Marangoni convection inside oil menisci surrounding central objects with different thermal conductivities. (a) Interfacial velocity along the oil-air interface for thermal conductivities of k = 0.3, 0.6, 1.14 and 21 W/(m·K), respectively. Outward interfacial oil flow is defined as positive. The substrate temperature is set as 283.15 K. Temperature distribution and flow field inside the oil meniscus surrounding a central object with k = 21 W/(m·K).*

## S6 Negative convection roll in an oil meniscus on a heated substrate

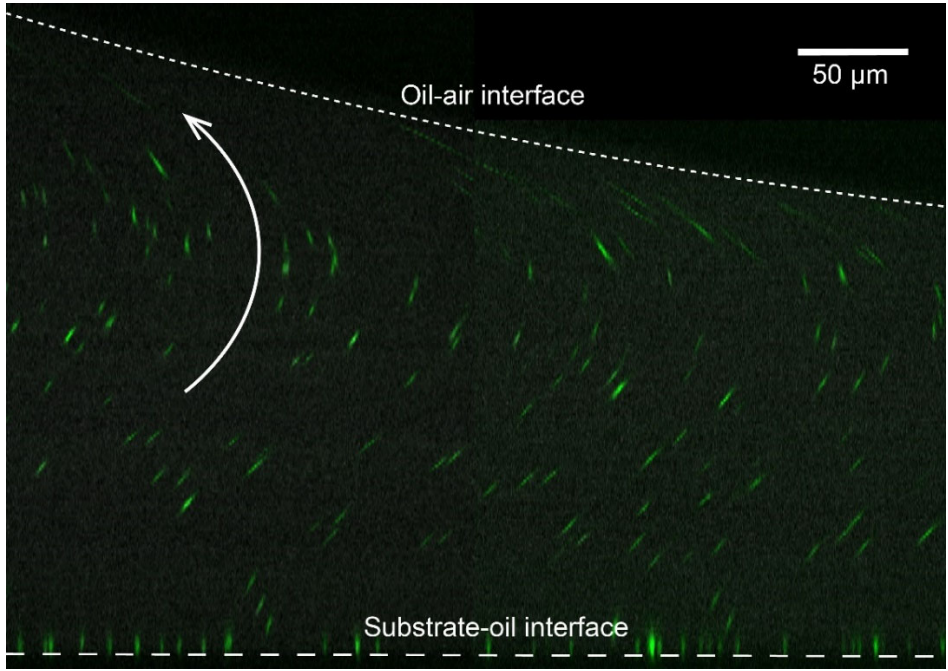

*Figure S6. Superimposed orthogonal images from a confocal microscopy z-stack scan of the convection flow in a 48-cP silicone oil meniscus surrounding a 780-μm glass sphere (to the left of this image) on a sapphire substrate held at 35 °C. The oil was dyed with Lumogen Rot F305 and labeled with a low concentration of 1-μm Fluoresbrite YG microspheres. The particle motion can be seen as streamlines, since both the particles and the scanning direction moved from the bottom to the top of the meniscus. One z-scan took approximately 62 milliseconds to complete.*

## S7 Estimation of (cos $\varphi_F$ – cos $\varphi_B$) for large supported and small floating droplets

The secondary meniscus (belonging to the moving droplet) is modified due to the presence of the curved primary oil meniscus, causing an angle difference (cos $\varphi_F$ − cos $\varphi_B$) and thereby the unbalanced capillary force, as shown in **Fig. S7(a)**. The angle difference for large supported droplets is experimentally measured from side-view during the droplet movement, as shown in **Fig. S7(b)**, and is (cos $\varphi_F$ − cos $\varphi_B$) ∼ $10^{-1}$. For small floating droplets, (cos $\varphi_F$ − cos $\varphi_B$) is related to the angle difference of the primary meniscus slope between points F and B (intersection of the droplet with the primary meniscus at the front and back, with $\beta_F$ and $\beta_B$, respectively), as illustrated in **Fig. S7(c)**. The angle difference (cos $\beta_F$ − cos $\beta_B$) is determined by the curvature of the primary meniscus and the droplet size. If the oil meniscus were replaced with a flat oil film, (cos $\varphi_F$ − cos $\varphi_B$) = (cos $\beta_F$ − cos $\beta_B$) = 0 for all droplets. Similarly, to small floating droplets, the primary meniscus appears nearly planar. We experimentally measured a meniscus profile and fit it with an exponential expression: $h(r) = 1487 \times e^{-0.004 \times r} + 4.62$ (see **Fig. S7(d)**). Suppose there is a small floating droplet ($d$ = 25 μm) at $r$ = 800 μm from the center of the sphere, we obtain cos $\varphi_F$ – cos $\varphi_B$ ≈ cos $\beta_F$ – cos $\beta_B$ ≈ 0.004.

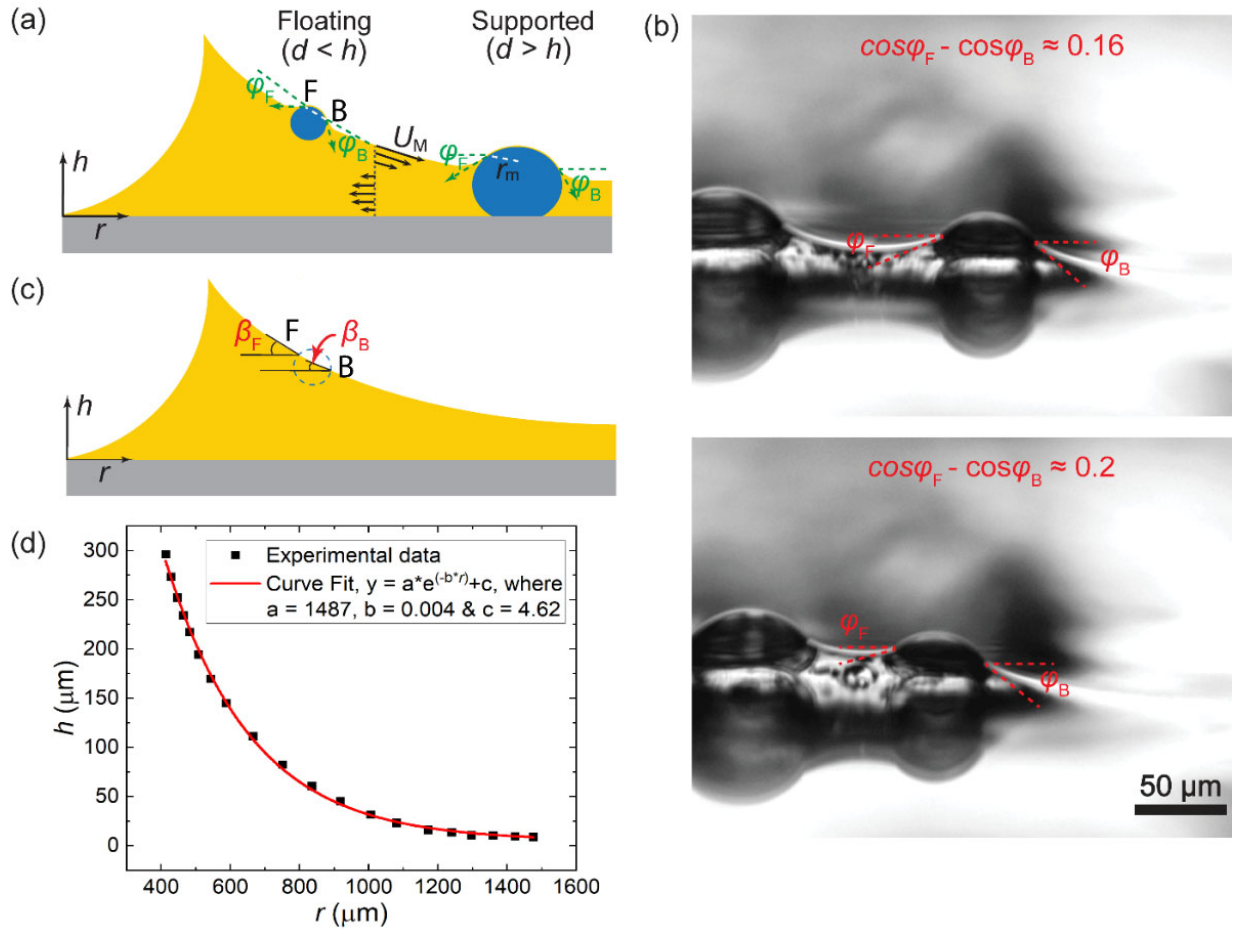

*Figure S7. Schematics of the angle difference (cos $\varphi_F$ – cos $\varphi_B$) for large supported and small floating droplets on an oil meniscus. (b) Estimation of the angle difference for a supported droplet during the movement based on side-view images. (c) Approximation of the angle difference for a floating droplet as the local meniscus slopes at points F and B. (d) Exponential fitting of the oil-air interfacial profile of an oil meniscus as experimentally measured by confocal fluorescence microscopy.*

# S8 Dynamics of droplets initially dispensed closer to the meniscus top

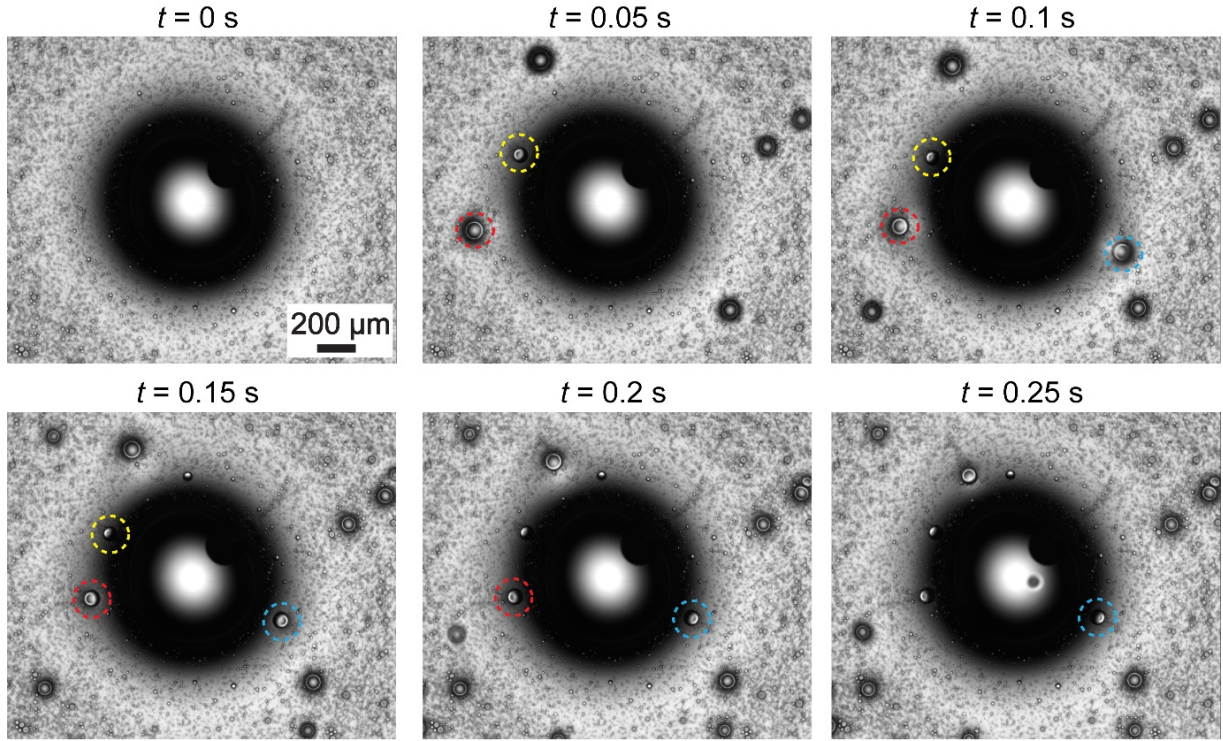

*Figure S8. Droplet movements when initially deposited closer the top of the meniscus. An ethylene-glycol droplet with a diameter of 650 um was pre-placed on a LIS with a layer of 19 cP silicone oil (initial thickness ~ 16 μm). Then, water droplets were added to the sample using a mist sprayer and humidifier. Select droplets, which are initially close to the apex of meniscus are highlighted in colorful dashed circles. These large supported droplets still ascend the meniscus and move towards the central droplet.*

# S9 Determination of fitting factor *f*

We repeated the simulation outlined in **Section S4** and varied the substrate temperature from 10 to 50°C, with all other parameters staying the same. **Figure S8** shows the temperature variation along the oil-air interface, $\Delta T_{\text{interface}}$, *versus* the temperature difference between substrate and environment, $T_s - T_{\text{env}}$ (*i.e.*, the vertical temperature difference). We define the fitting factor as $f = \Delta T_{\text{interface}} / (T_s - T_{\text{env}})$, which is related to the Biot number $Bi = h_c\sqrt{h_m^2 + L^2}/k_{\text{oil}}$ of the oil film, where $h_c$ is the convective heat transfer coefficient at the oil-air interface and $k_{\text{oil}}$ is the thermal conductivity of the oil film.

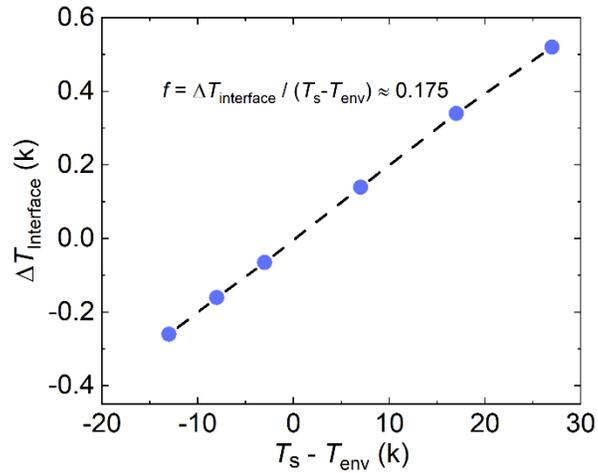

*Figure S9. Simulation results of the oil-air interfacial temperature difference along the meniscus arc length as a function of the substrate temperature relative to the environment.*